\documentclass[aps,preprint]{revtex4}
\usepackage{amsfonts}
\usepackage{xspace}
\usepackage{relsize}
\usepackage{amssymb}
\usepackage{amsmath}
\usepackage[font={footnotesize,it}]{caption}
\usepackage{latexsym}
\usepackage{changes}
\usepackage{hyperref}
\usepackage[capitalise]{cleveref}
\usepackage{lipsum}

\usepackage[caption = false]{subfig}
\usepackage{graphicx}
\usepackage{float}
\usepackage[font=small,labelfont=bf,
   justification=justified,
   format=plain]{caption} 
\hypersetup{
    colorlinks=true,
    linkcolor=blue,
    filecolor=magenta, citecolor=blue,    urlcolor=blue,
}
\begin{document}
%\title{Sections and Chapters}

\title{Reply to ``Comment on `Greybody radiation and quasinormal modes of Kerr-like black hole in
Bumblebee gravity model' ''}
\author{Sara Kanzi}
\email{sara.kanzi@emu.edu.tr}
\author{\.{I}zzet Sakall{\i}}
\email{izzet.sakalli@emu.edu.tr}
\affiliation{Department of Physics, Eastern Mediterranean
University, G. Magusa, 99628 North Cyprus, Mersin-10, Turkey.}


\begin{abstract}

In our recent study \big[Ref. \cite{DB}: Eur. Phys. J. C \textbf{81}, 501 (2021)\big] , the main goal was to reveal the Lorentz symmetry breaking (LSB) effect in the rotating black holes via  the greybody factor and quasinormal mode analyzes. Ding et al.'s \cite{er1}  Kerr-like black hole solution of the bumblebee gravity model used in our paper \cite{DB} has unquestionably proven by Maluf and Muniz \cite{com} that it is wrong. We now explain under which condition the greybody factor and quasinormal mode analyzes performed in \cite{DB} become valid. We re-make the calculations according to the condition in question and present the new results in this reply.
\end{abstract}
%\maketitle
\date{\today}
{
\let\clearpage\relax}
\maketitle
The comment \cite{com} on our paper \cite{DB} does not target the greybody factor (GF) and quasinormal mode (QNM) analyzes of the Kerr-like black hole in the bumblebee gravity model, but it is all about the error of the black hole solution obtained in \cite{er1}. In fact, we did the GF and QNM studies with high confidence, based on Ding et al.'s paper \cite{er1}, as the supposed new Kerr-like black hole solution has passed the serious peer-reviews of one of the most respected journals in the world. Similar to us, in the recent past, it is possible to see that there are many quality studies that have directly used that black hole solution (see for example \cite{ex1,ex2,ex3,ex4}).

On the other hand, after an in-depth analysis, we have also realized that the supposed Kerr-like black hole solution in bumblebee gravity model \cite{er1} is wrong, as highlighted by \cite{com}. Let us recall the metric of this wrong black hole solution \cite{er1}. In the Boyer-Lindquist coordinates, it was given by 
\begin{equation}
ds^{2}=-\left(1-\frac{2 M r}{\Sigma}\right) d t^{2}-\frac{4 M r a \sqrt{1+L} \sin ^{2} \theta}{\Sigma} d t d \varphi+\frac{\Sigma}{\Delta} d r^{2}+\Sigma d \theta^{2}+\frac{A \sin ^{2} \theta}{\Sigma} d \varphi^{2}, \label{iz1}   
\end{equation}
where $L=\varrho b^{2}$ represents the contribution coming from the Lorentz-violating effects and the functions $\Sigma$, $\Delta$, and $A$ are as follows
\begin{multline}
 \Sigma(r, \theta)=r^{2}+(1+L) a^{2} \cos ^{2} \theta, \quad \quad  \Delta(r)=\frac{r^{2}-2 M r}{1+L}+a^{2}, \\
 \quad A(r, \theta)=\left[r^{2}+(1+L) a^{2}\right]^{2}-\Delta(1+L)^{2} a^{2} \sin ^{2} \theta.   \label{iz2}    
\end{multline}

Meanwhile, $\varrho$ denotes a real coupling constant (with mass dimension $-1$ ) that controls the non-minimal gravity interaction to bumblebee field $B_{\mu}$ (having the mass dimension $1$) and $b^{2}$ is a real positive constant. The bumblebee field strength is defined by
\begin{equation}
B_{\mu \nu}=\partial_{\mu} B_{\nu}-\partial_{\nu} B_{\mu}. \label{iz3}     
\end{equation}
and it is fixed to $B_{\mu}=b_{\mu}$. Thus, the potential $V$, which drives the Lorentz symmetry breaking and takes a minimum at $B_{\mu} B^{\mu} \pm b^{2}=0$,
becomes zero and also its derivative vanishes: $V^{\prime}=0$. Someone else like us can check that the equations of motion for the bumblebee field (with $\kappa=8\pi$):
\begin{equation}
\nabla^{\mu} b_{\mu \nu}+\frac{L}{\kappa b^2} b^{\mu} R_{\mu \nu}=0, \label{iz4}    
\end{equation}
can not be satisfied with metric \eqref{iz1} and the following bumblebee field given in \cite{er1}
\begin{equation}
 b_{\mu}=\left(0, \sqrt{\frac{b^2 \Sigma}{\Delta}}, 0,0\right),   
\end{equation}
which yields $b_{\mu} b^{\mu}=b^{2}=C$ where $C$ is a positive constant. In particular, as stated in \cite{com}, the following field equations, which should normally result in zero, are failed:

\begin{multline}
\nabla^{\theta} b_{\theta r}+\frac{L}{\kappa b^2} b^{r} R_{r r}=\frac{a^{2}}{2 b \kappa\left(r^{2}+a^{2}(1+L) \cos ^{2} \theta\right)^{3}} \sqrt{\frac{(1+L)\left(r^{2}+a^{2}(1+L) \cos ^{2} \theta\right)}{r^{2}-2 M r+a^{2}(1+L)}} \times \\
{\left[a^{2}(1+L) \cos ^{2} \theta\left(3 L^{2}-4 b^{2} \kappa(1+L)+L^{2} \cos 2 \theta\right)+r^{2}\left(L^{2}-b^{2} \kappa(1+L)\right)(1+3 \cos 2 \theta)\right]}\neq 0, \label{iz5}
\end{multline}

and

\begin{multline}
\nabla^{r} b_{r \theta}+\frac{L}{\kappa b^2} b^{r} R_{r \theta}=\frac{a^{2} b \sin 2 \theta}{4\left(r^{2}+a^{2}(1+L) \cos ^{2} \theta\right)^{3}} \sqrt{\frac{(1+L)\left(r^{2}+a^{2}(1+L) \cos ^{2} \theta\right)}{r^{2}-2 M r+a^{2}(1+L)}} \times \\
{\left[r\left(a^{2}(1+L)(-5+\cos 2 \theta)+10 M r-4 r^{2}\right)-2 a^{2}(1+L) M \cos ^{2} \theta\right]}\neq 0. \label{iz6}
\end{multline}

In short, the alleged Kerr-like rotating black hole solution presented in \cite{er1} is not correct. Therefore, the spinning black hole solution remains an open question for the Einstein-bumblebee gravity theory. We have also understood that Newman-Janis algorithm \cite{NJA} does not work for such a nonlinear source (see \cite{non,Ding}). On the other hand, in the slowly rotating limit ($a^{2} \rightarrow 0$), all the field equations are fulfilled \cite{com}.

Even though metric \eqref{iz1} is incorrect, if the slow rotating limit ($a^{2} \rightarrow 0$) of the metric \eqref{iz1} is considered, we get 
\begin{equation}
d s^{2}	\approx-\left(1-\frac{2 M}{r}\right) d t^{2}-\frac{4 M \tilde{a} \sin ^{2} \theta}{r} d t d \varphi+\frac{(1+L) r}{r-2 M} d r^{2}+r^{2} d \theta^{2}+r^{2} \sin ^{2} \theta d \phi^{2}, \label{iz7}
\end{equation}

where $\tilde{a}=\sqrt{1+L}a$. Metric \eqref{iz7} is essentially nothing but the \textit{true slowly rotating black hole solution of Einstein-bumblebee gravity model} presented in article \cite{Ding} [see Eq. (3.15) therein] if we apply the following mapping: $\tilde{a} \rightarrow a$. Therefore, if we first take the slow rotation limit of all our computations of GFs and QNMs seen in our study  \cite{DB} and in the sequel apply the relevant mapping $\tilde{a} \rightarrow a$, we can correct all the results obtained from the wrong metric \cite{er1}. We would like to highlight that there is no problem in the methodology (independent of the consider metric) that we applied for finding the GFs and QNMs. In this context, under the slow rotation condition, one can get the following expressions, graphs, and tables for metric \eqref{iz7} with $\tilde{a} \rightarrow a$. 

\begin{figure}[h]
\centering
\includegraphics[width=10cm,height=8cm]{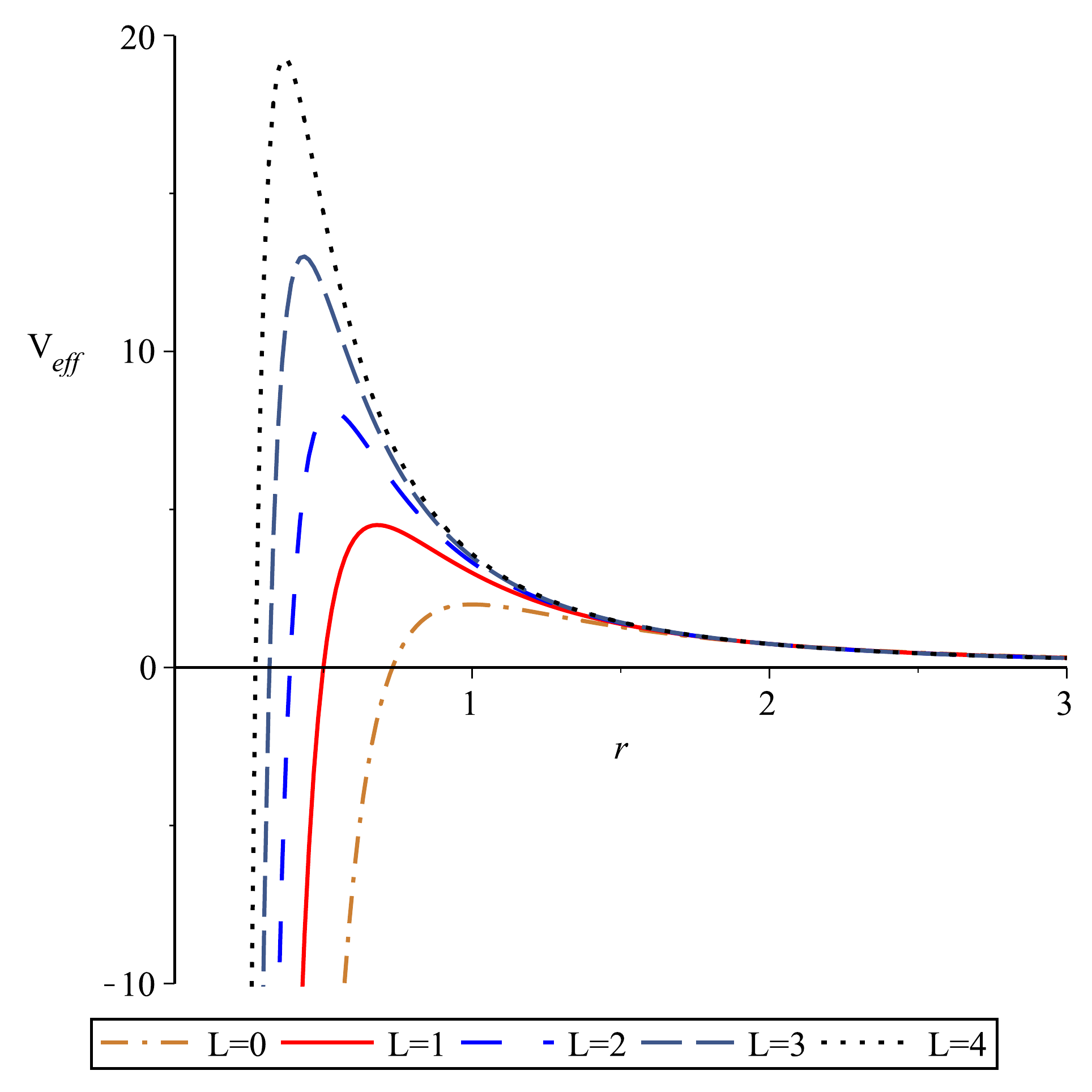}\caption{Plot of $V_{eff}$ versus $r$ for bosonic (spin-$0$) particles. The physical parameters are chosen as follows $M=m=1, \omega=15, a=0.1$, and $\lambda=2$.}%
\label{iz9}%
\end{figure}

\begin{figure}[h]
\centering
\includegraphics[width=10cm,height=8cm]{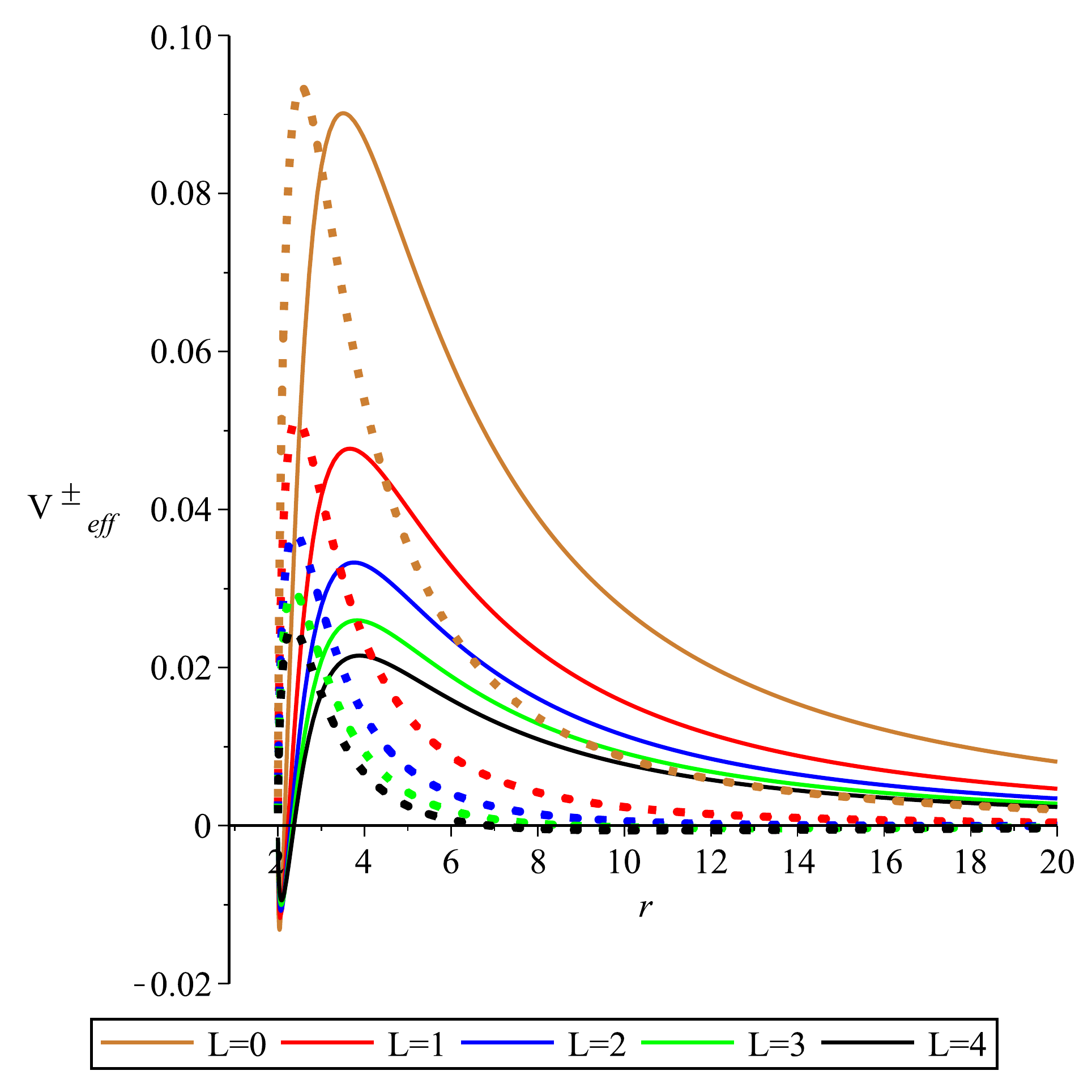}\caption{Plot of $V_{eff}^{\pm}$ versus $r$ for spin-$\pm\frac{1}{2}$ (up/down) particles. While the solid lines stand for the spin-up particles, the dots are for the spin-down particles. The physical parameters are chosen as follows: $M=1$, and $\lambda=-1.5$.}%
\label{iz11}%
\end{figure}

Effective potential for bosons \big[see Eq. (3.10) of Ref. \cite{DB}\big] becomes:  
\begin{equation}
V_{\text {eff }}=\frac{\left(1-\frac{2M}{r}\right)}{r^2}\left[\frac{2M}{r\left(1+L\right)}+\mu^2r^2+\lambda\right]+\frac{4Mm\omega a}{r^3},  \label{iz8}
\end{equation}
and its corresponding plots are depicted in Fig. \eqref{iz9}. 
Moreover, the fermionic effective potentials \big[see Eq. (4.29) of Ref. \cite{DB}\big] are recomputed as
\begin{equation}
V_{\text {eff }}^{\pm}=\frac{\lambda\left(1-\frac{2M}{r}\right)}{\sqrt{1+L}}\left[\frac{\lambda}{r^2\sqrt{1+L}}\pm\left(\frac{r-M}{r^2\sqrt{r^2-2Mr}}-\frac{2\sqrt{r^2-2Mr}}{r^3}\right)\right], \label{iz10}
\end{equation}
and their corresponding plots are illustrated in  Fig. \eqref{iz11}, which represents the behaviors of $V_{\text {eff }}^{\pm}$ with changing the LSB parameter for both spin-$\pm\frac{1}{2}$ particles. In general, the associated potentials' behaviors are changed significantly compared to those seen in \cite{DB}.  

GFs of the bosons based on metric \eqref{iz7} are recomputed as follows
\begin{equation}
\sigma_{l}\left(\omega\right)\geq\sec h^{2}\left\{ \frac{1}{2\omega} \int_{r_{h}}^{\infty}\left(\frac{2M}{r^3\sqrt{1+L}}+\frac{\lambda\sqrt{1+L}}{r^2}+\frac{4Mma\sqrt{1+L}}{r^2\left(r-2M\right)}\right)\right\}, \label{iz12}
\end{equation}
which yields
\begin{equation}
\sigma_{l}\left(\omega\right)\geq\sec h^{2}\left\{ \frac{1}{2\omega} \left(\frac{M}{r_{h}^3\sqrt{1+L}}+\frac{\lambda\sqrt{1+L}}{r_{h}}+4Mma\sqrt{1+L}\left(\frac{1}{2r_{h}^2}+\frac{M}{4r_{h}^4}\right)\right)\right\}.
\label{iz13}
\end{equation}
The plots that depict the LSB effect on the GFs \eqref{iz13} of the bosons are shown in Fig. \eqref{iz14}. During these numerical studies, it is observed \big(see also Fig. \eqref{iz14}\big) that at low $L$ values ($L=0..4$) while GFs increase with the increasing LSB effects, at high ($L>4$) values, the situation reverses and GFs decrease with the increasing LSB effects.
\begin{figure}[h]
\centering
\includegraphics[width=10cm,height=8cm]{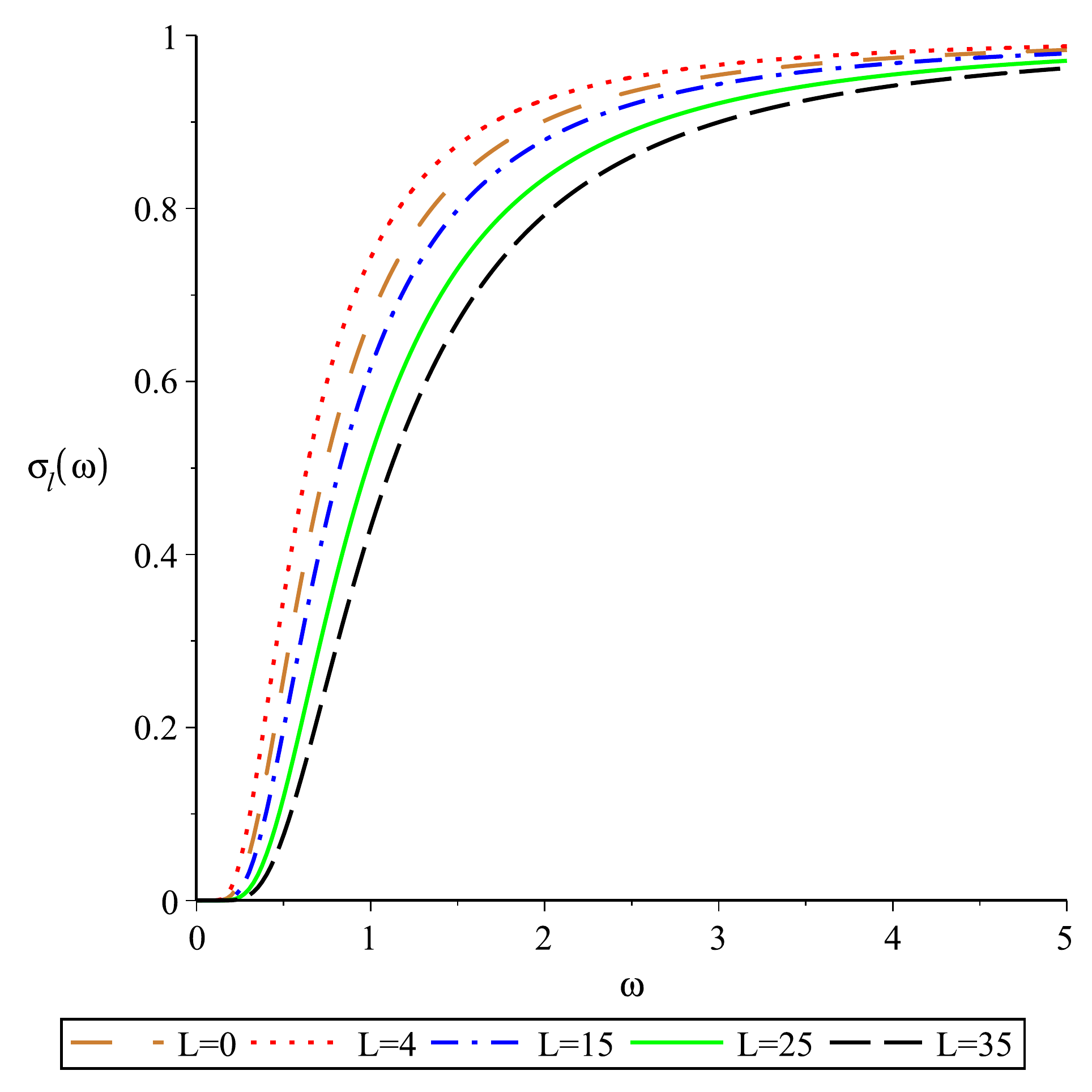}\caption{Plot of $\sigma_{l}\left(\omega\right)$ versus $\omega$ for spin-$0$ particles. The physical parameters are chosen as follows: $M=m=r=1, a=0.1$, and $\lambda=0$.}%
\label{iz14}%
\end{figure}

Furthermore, GFs of the fermions propagating in the spacetime \eqref{iz7} are obtained as follows 
\begin{equation}
\sigma_{l}\left(\omega\right)\geq\sec h^{2}\left\{ \frac{1}{2\omega} \int_{r_{h}}^{\infty}\left(\frac{\lambda^2}{r^2\sqrt{1+L}}\pm\lambda\frac{3M-r}{r^2\sqrt{r^2-2Mr}}\right)\right\},
\label{iz14}
\end{equation}
which results in
\begin{equation}
\sigma_{l}\left(\omega\right)\geq\sec h^{2}\left\{ \frac{1}{2\omega} \left(\frac{\lambda^2}{r_{h}\sqrt{1+L}}\pm\lambda\left[\frac{-1}{r_h}+\frac{M}{r_{h}^2}+\frac{3M^2}{6r_{h}^2}+\frac{9M^3}{8r_{h}^4}\right]\right)\right\}.
\label{sk9}
\end{equation}
\begin{figure}[h]
\centering
\includegraphics[width=10cm,height=8cm]{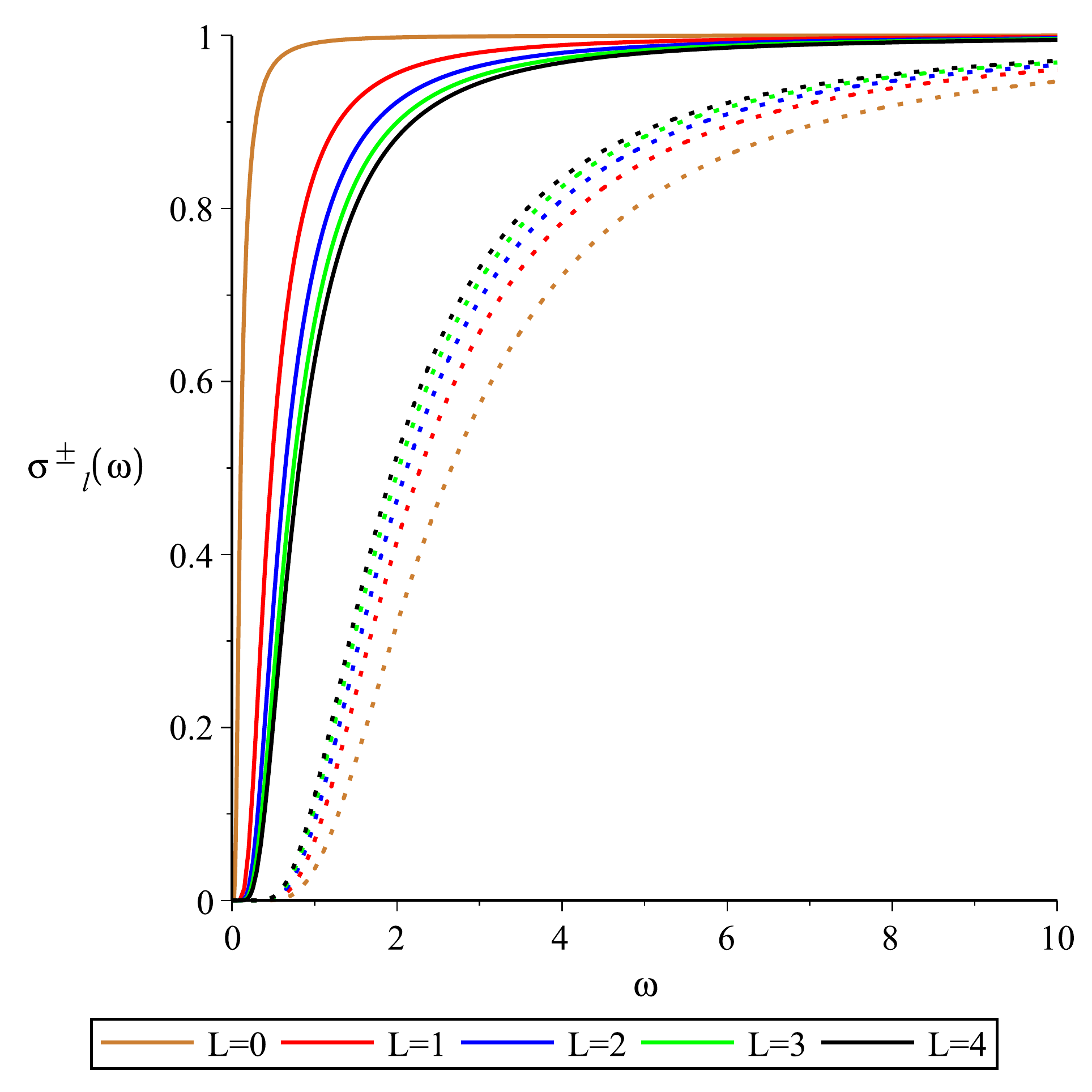}\caption{Plot of $\sigma_{l}^{\pm}\left(\omega\right)$ versus $\omega$ for spin-$\pm\frac{1}{2}$ particles. While the solid lines stand for the spin-up particles, the dots are for the spin-down particles. The physical parameters are chosen as follows: $M=r=1$ and $\lambda=-1.5$.}%
\label{iz15}%
\end{figure}

Finally, we present the new bosonic and fermionic QNMs computed for the metric  \eqref{iz7}, whose detailed methodology was given in section (6) of our article \cite{DB}, in Tables \eqref{iz15} and \eqref{iz16}, respectively. The bosonic QNMs are tabulated in Table \eqref{iz15} in which the physical parameters are chosen as follows: the angular momentum $l=2$, the magnetic quantum number $m=0$, and overtone quantum number $n=0$ (for the higher overtone numbers, the bosonic QNMs exhibit similar behaviors with the fundamental mode, $n=0$). The revealed information obtained from the both Tables are that while both the oscillation and damping rates of bosonic QNMs increase with the LSB parameter, the fermionic QNMs show opposite behavior and their oscillation and damping rates decrease with the increasing LSB effect. It is also worth noting that one of the most striking results of the slow rotation approximation is that fermionic GFs and QNMs are independent of the spin parameter $a$, whereas the bosonic GFs and QNMs have this dependence.

\begin{table}
  \centering
    \begin{tabular}{ |c|c| }
\hline
$L$ & $\omega_{bosons}$ \\
\hline
 0 &  0.6193445868-0.4259671652i\\
 1 &  0.7688968549-0.6024536531i \\
1.1 & 0.7791058962-0.6159336514i \\
1.2 & 0.7887835943-0.6289091139i \\
1.3 & 0.7979767859-0.6414186083i \\
1.4 & 0.8067264683-0.6534964325i \\
1.5 & 0.8075365232-0.6587911664i \\
1.6 & 0.8230350287-0.6764758541i \\
1.7 & 0.8306539984-0.6874294407i \\
1.8 & 0.8379506594-0.6980559713i \\
1.9 & 0.8449475817-0.7083755957i\\
2   & 0.8516649690-0.7184065252i\\
\hline 
    \end{tabular}
  \caption{Table 1: QNMs of scalar waves
in the slowly rotating Kerr-like black hole
spacetime.} \label{iz15}
\end{table}

\begin{table}
  \centering
    \begin{tabular}{ |c|c|c|c| }
\hline
$l$ & $n$ & $L$ & $\omega_{fermions}$ \\
\hline
 1 & 0 &  0  &  0.1615755163-0.1986761458i\\
   &   & 0.1 &  0.1057644220-0.1612179485i \\
   &   & 0.2 &  0.0530479984-0.1300388664i \\
   &   & 0.3 &  0.0019930161-0.1113914550i \\
   & 1 &  0  &  0.2408767549+0.5575170642i \\
   &   & 0.1 &  0.3028756279+0.5575297813i \\
   &   & 0.2 &  0.3477807164+0.5531523256i \\
   &   & 0.3 &  0.3780138537+0.5433679210i \\
2  & 0 &  0  &  1.4891900030-0.8296137229i \\
   &   & 0.1 &  1.0377655280-0.3211236139i \\
   &   & 0.2 &  0.8543487310-0.3189534898i\\
   &   & 0.3 &  0.7108515298-0.3159381339i\\
   &   & 0.4 &  0.5961995167-0.3122465225i\\
\hline 
    \end{tabular}
  \caption{Table 2: QNMs of Dirac waves in the slowly rotating Kerr-like black hole
spacetime.} \label{iz16}
\end{table}

In conclusion, the above consideration shows that although the metric solution provided by Ding et al \cite{er1} and which we also used it in our recent study \cite{DB} \big(like the others \cite{ex1,ex2,ex3,ex4}\big) is incorrect, the GF and QNM results obtained in \cite{DB}  can be adopted for metric \eqref{iz7} (with $\tilde{a} \rightarrow a$), which represents the slowly rotating Kerr-like black hole solution to the Einstein-bumblebee gravity equations \cite{non}. Besides, we want to emphasize that our GFs and QNMs analyzes in general do not strongly depend on the type of metrics. Namely, the considered metric only affects the results obtained not the algorithm of finding the GFs and QNMs. 

In summary, we have corrected our GF and QNM analyzes, which are the main subjects of our previous study \cite{DB}, with this reply since the metric \cite{er1} used in \cite{DB} was incorrect.  

\section*{ Acknowledgements}

The authors are grateful to the Editor for giving a chance to them for correcting and improving their paper \cite{DB}.

\end{document}